\documentclass[prl,aps,twocolumn,floatfix,nofootinbib,showpacs,superscriptaddress]{revtex4-1}

\usepackage{float,mathtools,mathrsfs,amsmath,amsfonts,amssymb,slashed,bm,graphicx,color}

\begin{document}

\title{Emergence of Quantum Theory and Minkowski Spacetime}
\author{Si-xue Qin}
\email{sqin@cqu.edu.cn}
\affiliation{Department of Physics and Chongqing Key Laboratory for Strongly Coupled Physics, Chongqing University, Chongqing 401331, P. R. China.}

\begin{abstract}
The state-of-the-art physics consists of two irreconcilable branches, i.e., the quantum theory and the general relativity, which work well in their own territories, independently. However, what are quantum and spacetime after all? The key question was never addressed, satisfactorily. In this work, we describe a possibility to reformulate the quantum theory in the Minkowski spacetime from the viewpoint of classic physics in the Euclidean spacetime, i.e., classic mechanics and stochastic process theory. We show that quantum theory and Minkowski spacetime may connect with each other and emerge from a single fundamental entity.
\end{abstract}


\date{\today}

\maketitle

\noindent \textbf{Introduction~}
Quantum theory was born from the simple idea that energy can be absorbed or released only in tiny, differential, discrete packets. It has been developed for more than one hundred years. Nowadays, it is generally believed that quantum is the key feature of the Nature's fundamental laws. In modern physics, quantum has been generally summarized as three basic assumptions or principles: 1) Physical states are represented by rays in Hilbert space; 2) Observables are represented by Hermitian operators; 3) Experiments can only observe a series of eigenvalues with possibilities. These principles are the short answers of the following three primitive and intuitive questions: 1) How to describe a physical system; 2) How to express an observation; 3) How to explain the experimental results? Based on the answers, the most successful quantum theories are quantum mechanics and quantum field theory. The former has been widely used in atomic, molecular, and condensed matter physics \cite{Dirac1958}; and the latter provides a unique way to reconcile quantum mechanics with special relativity, and has inconceivably unified three of four fundamental forces in the Nature \cite{Weinberg1995}. 

The foundation of quantum mechanics opened a new era of physics. Since then, the physics was divided into classical physics and quantum physics. In general, the Planck constant $\hbar$ is a signature of quantum physics. If it is set to a limit: $\hbar \to 0$, then quantum physics degenerates to classical physics. Almost, ``classical" always refers to outdated and approximate, while ``quantum" to modern and accurate. In such a context, for a classical theory, it is always eager to develop the corresponding quantum version. Since last century, a quantum theory of everything has been an ultimate dream of physicists. Unfortunately, after efforts of several generations, the dream is still far-reaching. The state-of-the-art physics consists of two irreconcilable branches: the quantum theory and the general relativity govern the micro-scale and macro-scale worlds, respectively. 

Of course, the step to pursue the ultimate dream never stoped. In the path, some crucial and deep questions are raised. On the one hand, physicists can use quantum theory with perfect confidence. However, it is just a black box. What is quantum after all? In fact, the basic principles of quantum theory are quite abstract. As the famous Richard Feynman observed ``I think I can safely say that nobody really understands quantum mechanics". Up to now, there is still no complete consensus as to what the quantum theory tells us about reality, or even whether ``reality" exists at all. On the other hand, the key insight of general relativity is to identify gravity as an emergent effect of spacetime dynamics. However, what is spacetime itself? Without a knowledge of the essence of spacetime, it is impossible to achieve a satisfactory theory of quantum gravity. Recently, it is becoming popular that spacetime and gravity must emerge from something more fundamental. 

Now we arrive at a point to address the key questions about quantum and spacetime. Is there a possibility that the two questions are related together and can be addressed by a single answer, if the so-called quantum is not like what we have been brainwashed to accept, and merely reflects our ignorance on the reality? Historically, the attempt has been made, which is referred to local hidden-variable theories. What then happened is well-known. The famous Bell's theorem concludes: No physical theory of local hidden variables can ever reproduce all of the predictions of quantum mechanics. However, the exact nature of the assumptions required to prove a Bell-type constraint on correlations is still in debate. Moreover, the experimental improvements of ``closing loopholes in Bell tests" are also in process. In this work, we try to provide a solution for the key questions based on a couple of conclusions which are quite generally true in classical physics.

~\par\noindent \textbf{Hamilton's principle}
For a classical mechanical system, the motion can be described by a collection of time-dependent general coordinates: $q^i(t)$. The Hamilton's principle states that the system must ``choose" the trajectory $q^i(t)$ to make a stationary action functional:
\begin{align}
	\mathcal{S}[q^i(t)] = \int_{q^i(t_0)}^{q^i(t)} L (q^i,\dot{q}^i,t') dt'\,,
\end{align}
where the Lagrangian $L (q^i,\dot{q}^i,t)$ is defined by the general coordinates and velocities. The stationary condition gives the equation of motion of the system, i.e., the Euler-Lagrange equation:
\begin{align}
	\frac{d}{dt} \frac{\partial L}{\partial \dot{q}^i} - \frac{\partial L }{\partial q^i} = 0 \,.
\end{align}
At this point, we only use the terminology ``time" to parameterize the sequences of general coordinates. The true nature of time is to label states of the system in the past and the future which we usually call. Accordingly, we introduce no assumption on the structure of the entirety of ``spacetime". 

In order to make the above conclusion more useful, we need further steps. Making a Legendre transform and introducing the momentum as
\begin{align}
	p^i = \frac{\partial L }{\partial \dot{q}^i}\,,
\end{align}
we can define the Hamiltonian: $H(p^i,q^i,t)=p_i\dot{q}^i - L$ and translate the Euler-Lagrange equation into the Hamilton equations as
\begin{align}
	\frac{d q^i}{dt} = \frac{\partial H}{\partial p^i},\quad \frac{d p^i}{dt} = -\frac{\partial H}{\partial q^i}\,.
\end{align}
The above equations guarantee that the Hamiltonian has the conservation law as
\begin{align}
	\frac{d H}{dt} = \frac{\partial H}{\partial p^i}\dot{p}^i + \frac{\partial H}{\partial q^i}\dot{q}^i + \frac{\partial H}{\partial t} = \frac{\partial H}{\partial t}\,,
\end{align}
which means that the Hamiltonian is constant if it has no explicit time-dependence for a close system.

Now thinking of the abstract space $X = (q^i,t)$ and introducing two vectors:
\begin{align}
	P := (p^i, -H),\quad Q := (q^i, H)\,,
\end{align}
we can analyze the following integral as
\begin{align}
	{S} = \int P_i dX^i = p_iq^i\big|_{t_0}^{t} - \int Q_i dX^i\,,
\end{align}
where the integral interval of time is $[t_0,t]$ and $[q_0^i,q^i]$. Then, the necessary and sufficient condition for the path independence of the integral is written as:
\begin{align}
	\frac{\partial P^i}{\partial X^j} = \frac{\partial P^j}{\partial X^i},\quad \frac{\partial Q^i}{\partial X^j} = \frac{\partial Q^j}{\partial X^i}\,,
\end{align}
which is nothing else but Hamilton equations with $q^i$ and $p^i$ as independent variables. Therefore, the integral ${S}$ is a function in the space of $X$. And the action functional is called on-shell if and only if
\begin{align}
	\mathcal{S}[q^i(t)] = S(q^i_0,t_0; q^i,t)\,.
\end{align}

At last, we can have the total differential as
\begin{align}
	dS = \frac{\partial S}{\partial q^i} dq^i + \frac{\partial S}{\partial t} dt = p_i dx^i - H dt\,,
\end{align}
which arrives at the Hamilton-Jacobi equation \cite{Landau1971}:
\begin{align}
	\frac{\partial S}{\partial t} = - H\left(q^i,\frac{\partial S}{\partial q^i},t\right)\,.
\end{align}
For a close system, whose conserved Hamiltonian is $E$, we have the Hamilton-Jacobi equation as 
\begin{align}
	\frac{\partial S}{\partial t} = - E \,.
\end{align}

~\par\noindent \textbf{Random noise process}
If a system involves two types of dynamics with vastly different time scales, e.g., Brownian particles, it could be described as a random noise process. In general, this kind of motion is called Brownian motion, which is governed by the so-called Langevin equation:
\begin{align}
	\frac{d x}{dt} = -\beta(x) + f(t)\,,
\end{align}
where $x(t)$ is the general position of the Brownian particle. The key point of the Langevin equation is that the general position can be affected by two factors: the slow or static potential $\beta(x)$ and the fast or random force $f(t)$. In such a framework, the general position $x(t)$ is a time-dependent random variable, and thus we can only describe its probability distribution instead of predicting its exact values.

Now we introduce the conditional probability or Green function $\mathcal{P}(x,t|x_0,t_0)$, which is the probability density for the particle locating at the position $x_0$ when the time was $t_0$ to travel to the new position $x$ at the time $t$. Thus, the density satisfies the initial condition as
\begin{align}
	\mathcal{P}(x,t_0|x_0,t_0) = \delta(x-x_0)\,.
\end{align}
Again, ``time" is used to label events of the process. Assuming the random process is uncorrelated, i.e., Markov process, we have the so-called Smoluchowski equation as
\begin{align}
	\mathcal{P}(x,t|x_0,t_0) = \int_{-\infty}^{+\infty} &\mathcal{P}(x,t|x',t')\notag\\
	&\times \mathcal{P}(x',t'|x_0,t_0)\, dx'\,,
\end{align}
where an intermediate time is chosen as $t_0<t'<t$. 

Then, the time evolution of $\mathcal{P}(x,t|x_0,t_0)$ can be calculated as following
\begin{align}
	\frac{\partial \mathcal{P}}{\partial t} = \lim_{\Delta t\to 0} \frac{\mathcal{P}(x,t+\Delta t|x_0,t_0) - \mathcal{P}(x,t|x_0,t_0)}{\Delta t} \,,\label{eq:time_diff}
\end{align}
where the evolved probability density is written as
\begin{align}
	\mathcal{P}(x,t+\Delta t|x_0,t_0) = \int_{-\infty}^{+\infty} &\mathcal{P}(x,t+\Delta t|x',t)\notag\\
	&\times\mathcal{P}(x',t|x_0,t_0)\, dx' \,,\label{eq:evolved_prob}
\end{align}
with a proper intermediate time: $t_0<t<t+\Delta t$. Due to the initial condition, we have the limit for the integral:
\begin{align}
	\lim_{\Delta t\to 0} \mathcal{P}(x,t+\Delta t|x',t) = \delta(x-x')\,.
\end{align}
This mean that, for a small $\Delta t$, the probability density $\mathcal{P}(x,t+\Delta t|x',t)$ is a slightly-deformed $\delta$ function:
\begin{align}
	\mathcal{P}(x,t+\Delta t|x',t) \approx \delta(x-x') + f(x,x') \Delta t + \cdots\,.
\end{align}
Note that $f(x,x')$ may be constructed as the following expansion:
\begin{align}
	f(x,x') = \sum_{i=0}^{\infty} a_i(x') \delta^{(i)}(x-x')\,.
\end{align}
Once the coefficients $a_i$ are determined, one can plug the expansion into Eqs. \eqref{eq:time_diff} and \eqref{eq:evolved_prob} to obtain the evolution equation for $\partial\mathcal{P}/\partial t$.

The first observation is that the leading term of the expansion vanishes due to the probability conservation:
\begin{align}
	\int_{-\infty}^{+\infty} \mathcal{P}(x,t+\Delta t|x',t)\, dx = 1 + a_0(x')\Delta t = 1 \,.
\end{align}
In order to determine the rest of coefficients, we can analyze the probability density $\mathcal{P}(x,t+\Delta t|x',t)$ from two points of view. On the one hand, we can compute a series of statistical averages at the time $t+\Delta t$ as follows:
\begin{align}
	\overline{(x-x')^n} = \int_{-\infty}^{+\infty} (x-x')^n \mathcal{P}(x,t+\Delta t|x',t)\, dx\,,
\end{align}
which is also called the $n$-order moment. Plugging the expansion into the above equation and using the integral property of the delta function derivative, we can connect the coefficients with the moments order by order, e.g.,
\begin{align}
	\overline{(x-x')} = a_1(x')\Delta t, \quad \overline{(x-x')^2} = a_2(x')\Delta t\,.
\end{align}
On the other hand, we can directly compute the random variable $x$ by integrating the Langevin equation on a short time interval as
\begin{align}
	x-x' \approx -\beta(x')\Delta t + \int_t^{t+\Delta t} f(t')\, dt'\,.
\end{align}
Accordingly, we have the first-order moment as
\begin{align}
	\overline{(x-x')} = -\beta(x')\Delta t + \int_t^{t+\Delta t} \overline{f(t')}\, dt'\,.
\end{align}
Since the random force has a zero mean: $\overline{f(t')} = 0$, we can obtain the first-order coefficient as
\begin{align}
	a_1(x') = -\beta(x')\,.
\end{align}
Similarly, we can have the second-order moment as
\begin{align}
	\overline{(x-x')^2} = \beta^2(x')(\Delta t)^2 + \iint_t^{t+\Delta t} \overline{f(t')f(t'')}\, dt' dt''\,.\notag\\
\end{align}
Under the Markov assumption, the random force $f(t)$ is delta-function correlated as
\begin{align}
	\overline{f(t')f(t'')} = D \delta(t'-t'')
\end{align}
Keeping the first order of $\Delta t$, we then obtain
\begin{align}
	a_2(x') = D\,.
\end{align}
Further, it can be verified that the higher-order moments has no linear term of $\Delta t$, which means that
\begin{align}
	 a_{i}(x') = 0, \quad i\ge 3\,.
\end{align}
Up to now, the expansion has been completely determined by the features of the Langevin equation. At last, we obtain the so-called Fokker-Planck equation as \cite{Feller1950, Siegman1979}
\begin{align}
	\frac{\partial \mathcal{P}}{\partial t} = \frac{\partial}{\partial x} \left[ \beta(x) + D \frac{\partial}{\partial x} \right] \mathcal{P}\,.
\end{align}

~\par\noindent \textbf{Possibility space}
The spacetime is the place where all objects exist. Naively, it is imagined as a grid of continuous fabric along which objects can move. At the current stage, the most important insights on the spacetime can be summarized as two aspects: 1) The spacetime itself can have a dynamics which is governed by the general relativity; 2) The spacetime may have a fundamental structure at the Planck scale. Without further information on the spacetime, we can safely conclude that any system which we are interested in is inevitably perturbed by the spacetime as
\begin{align}
	\widetilde{H}(q^i,p^i,t) = H(q^i,p^i,t) + h(q^i,p^i,t)\,,
\end{align}
where $H$ is the Hamiltonian of the system. The perturbation sources, for example, may be gravitational waves which can propagate in the void or spacetime turbulences which the unknown structure and matter generates (dark energy/matter). In general, the perturbation $h$ could be a very complicated time-dependent function. If it has a characteristic time scale which is comparable to that of the Hamiltonian $H$, one may measure its possible affects and pin down its explicit expression. However, what if the perturbation varies at an extremely high frequency which present experiments cannot reach? In such a circumstance, the primary question is not what the exact form of $h$ looks like. Instead, we should ask if the perturbation can result in any visible consequences, i.e., how the dynamics of the perturbed system is rendered?

In the following discussion, we should keep in mind that there are two different characteristic time scales for the system and the perturbation, denoted by $\tau_s$ and $\tau_p$, respectively. Considering the hierarchy of the time scales, we have an interesting relationship between the system and the perturbation: From the system's viewpoint, the perturbation $h$ behaves as a random variable; from the perturbation's viewpoint, the system's Hamiltonian $H$ keeps constant. Accordingly, we can find an intermediate time interval $\Delta t$ as
\begin{align}
	\tau_p \ll \Delta t \ll \tau_s \,.\label{eq:timescales}
\end{align}
During such a time interval, the total Hamiltonian of the perturbed system can be rewritten as
\begin{align}
	\widetilde{H}(t) = E + \epsilon(t)\,,
\end{align}
where $E$ is the constant energy of the system and $\epsilon(t)$ is the random variable of the perturbation. 

We first assume that the perturbation is space-uncorrelated. This is sensible because a high frequency means a short characteristic space scale which must be much smaller the size of the system. Here, we have a thermodynamical analogy: The system can be considered as an equilibrium macroscopic body which can exchange energy with an environment created by the perturbation. Since the internal energy is additive, its fluctuation is proportional to the system size and thus itself. Then, the statistical variance of $\epsilon$ should be proportional to the mean value of the energy as
\begin{align}
	\overline{\Delta\epsilon^2} = \hbar E\,,
\end{align} 
where $\hbar$ is a small constant which characterizes the weak perturbation. Note that $\hbar$ does not have to be the Planck constant yet. Second, we can also assume that the perturbation is time-uncorrelated. Combining with the variance, we have the delta-function correlation function as
\begin{align}
	\overline{\epsilon(t)\epsilon(t')} = \hbar E\, \delta(t-t') \,.
\end{align}

Now plugging the total Hamiltonian $\widetilde{H}$ into the Hamilton-Jacobi equation, we can have
\begin{align}
	\frac{\partial S}{\partial t} = - E - \epsilon(t)\,,
\end{align}
which has the exact form of the general Langevin equation with the action $S$ working as a random variable. Accordingly, we obtain that the probability distribution of the action satisfies the Fokker-Planck equation as
\begin{align}
	\frac{\partial \mathcal{P}(S,t)}{\partial t} = \frac{\partial}{\partial S} \left[ E + \hbar E \frac{\partial}{\partial S} \right] \mathcal{P}(S,t) \,.
\end{align}
In terms of thermodynamics, we can consider the macroscopic evolution of the system as a quasi-static process, which happens slowly enough for the system to remain equilibrium against the perturbation, due to the hierarchy of time scales, i.e., Eq. \eqref{eq:timescales}. Therefore, the observed distribution must satisfy the stationary condition as
\begin{align}
	\frac{\partial \mathcal{P}(S,t)}{\partial t} = 0\,,
\end{align}
which has the following solution
\begin{align}
	\mathcal{P}(S) \propto e^{-S/\hbar}\,.
\end{align}

For the original unperturbed system, the action can only have an on-shell value, which corresponds to a unique trajectory. Due to the perturbation, the action becomes a random variable, which we also call the off-shell action. Thus, any trajectories $q^i(t)$ could be possible with probability densities as
\begin{align}
	\mathcal{P}[q^i(t)] \propto e^{-S[q^i(t)]/\hbar}\,,
\end{align}
where $S[q^i(t)]$ is the corresponding off-shell action functional. In the probabilistic terminology, if defining the collection of all trajectory configures of $q^i(t)$ as a set $\Omega$, we have a probability space with a positive measure as
\begin{align}
	d\mu = \frac{e^{-S[q^i(t)]/\hbar}}{\mathcal{Z}} \mathcal{D}q^i(t),~~ \int_\Omega d\mu = \mu(\Omega) = 1 \,,
\end{align}
where $\mathcal{D}q^i(t)$ is the measure of functional integral, and $\mathcal{Z}$ is the normalization constant to ensure the the probability conservation.

~\par\noindent \textbf{Quantum and spacetime}
At this point, it is time to clarify the true meaning of the time. At the first place, we have emphasized that the time $t$ is only used to label coordinates or events. Namely, the time $t$ does not have to be identical to the common definition. Here the point is that the time $t$ labels events of a quasi-static process. The second law of thermodynamics suggests that the quasi-static process is reversible only if the dynamics of the system is infinitely slow, otherwise it must be irreversible. That is to say, the labeled events must be ordered. Accordingly, the most important essence of time can emerge: Time is the order in which theories explain events. Therefore, any parameter that increases monotonically in that order can be a useful time coordinate.

Next we need to explain the relationship between space and time. For this purpose, we would like to use the framework of fields, which is able to be compatible with the possible structure of spacetime. Without losing any generality, we will adopt scalar fields for discussion. In a field theory, both space and time are used to label degrees of freedom of fields denoted by $\phi(t,\bm{x})$. In general, the fields are functions of space and time. However, since the perturbed fields can have any configurations, some of which may be badly behaved, the mathematical definition of fields must be generalized. Here comes the theory of distribution, which describes an object in terms of the way it behaves when integrated against a function as
\begin{align}
	\phi(f) := \int f(x) \phi(x) dx\,,
\end{align}
where $f$ is a well-behaved test function. In short, a distribution is defined by how it acts on test functions to give a number. For physical applications, we need a subclass of distributions, which is called tempered distributions, to preform Fourier transforms. Now, we can define a state of the field in its probability space as
\begin{align}
	|\phi\rangle := \sum_i c_i\, e^{i\phi(f_i)} \,,
\end{align}
where $c_i$ are complex numbers and $f_i$ are test functions. For different states, we have different $c_i$ and $f_i$. In order to express observables, we need to define a difference between states. Mathematically, the difference is called an inner product, and the probability space with inner products is a Hilbert space. For the definition of inner products, we have many choices, among which no one has a theoretical privilege. The one that can explain the Nature is expressed as
\begin{align}
	\langle \phi' | \phi \rangle := \sum_{i i'} {c'_i}^* c_i\, M_{ii'},~~ M_{ii'} = \int_\Omega e^{i \phi (f_i - \Theta f'_{i'})} d\mu \,,
\end{align}
where $\Theta$ is the time reflection operator defined as
\begin{align}
	\Theta f(t,\bm{x}) := f(-t,\bm{x}) \,,
\end{align}
and $M_{ii'}$ is a positive definite matrix, which is called reflection positivity. So, we have finished the construction of the physical Hilbert space.

Now, the structure of spacetime must preserve inner products in the physical Hilbert space, which implies a collections of symmetry transformations. Mathematically, these transformations can form the Euclidean group. It is worth noticing that, due to the ordering labeled by the time, the time translations can only form a semigroup. The spacetime with Euclidean covariance is called Euclidean spacetime. Combining with the reflection positivity, one can prove that a random field in the Euclidean spacetime has the same dynamics as a quantum field in the Minkowski spacetime. The correspondences includes two key aspects: 1) The time has to be continued to the imaginary axis by substituting $t\to it$; 2) The probability densities have to be translated to probability amplitudes as \cite{Osterwalder1973,Klein1983,Glimm1987}
\begin{align}
	\mathcal{P}[\phi] = \frac{1}{\mathcal{Z}[\phi]} e^{iS[\phi]/\hbar},~~ \mathcal{Z}[\phi] = \int \mathcal{D}\phi \, e^{iS[\phi]/\hbar}\,.
\end{align}
Conclusively, we arrived at the point where quantum theory and Minkowski spacetime emerge.

~\par\noindent \textbf{Conclusion}
In this work, we have reformulated the quantum theory in the Minkowski space from the viewpoint of classic physics. We first derived the Hamilton-Jacobi equation for the action function using the Hamilton's principle of classic mechanics, and then the Fokker-Planck equation for the general Langevin equation by the statistic theory of random noise process. Assuming that a high frequency perturbation exists in the spacetime, we considered the action function of the perturbed system as a random variable and the Hamilton-Jacobi equation as a special Langevin equation, then obtained the probability densities of off-shell trajectories by using a equilibrium thermodynamical analogy. Accordingly, we constructed the Euclidean random field theory, and translated it to the Minkowski quantum field theory from the reflection positivity and the Euclidean covariance of the physical Hilbert space. At last, we concluded that quantum theory and Minkowski spacetime could emerge from the classic level. Intuitively, we could imagine: If the spacetime is likened to water, then the general relativity explains the collective laws of water flows and the quantum theory describes the Brownian motions of aquatic objects.

\section{Acknowledgements}
\noindent The work was supported by the National Natural Science Foundation of China under Contract No. 11805024.

\end{document}